\documentclass[journal]{IEEEtran}

\usepackage[OT1]{fontenc} 
\usepackage[numbers,sort&compress]{natbib}
\usepackage[cmex10]{amsmath}
\usepackage{amssymb}
\usepackage{bm}
\usepackage{braket}
\usepackage{graphicx}
\usepackage{color}
\usepackage[caption=false,font=footnotesize]{subfig} 
\usepackage{tabularx}
\usepackage{arydshln}
\usepackage{mathtools}
\usepackage{multirow}
\usepackage{algorithm,algorithmic}
\usepackage{url}
\usepackage{listings}
\usepackage{comment}
\usepackage{CJKutf8}

\def\CN{\mathcal{CN}}
\def\I{\mathbf{I}}

\def\H{\mathbf{H}}

\def\s{\mathbf{s}}

\def\A{\mathbf{A}}

\def\b{\mathbf{b}}

\def\y{\mathbf{y}}
\def\U{\mathbf{U}}
\def\C{\mathbb{C}}

\def\x{\mathbf{x}}
\def\E{\mathbb{E}}

\def\W{\mathbf{W}}
\def\F{\mathrm{F}}
\def\e{\mathrm{H}}

\def\G{\mathcal{G}}

\def\P{\mathbf{P}}

\def\Q{\mathbf{Q}}

\def\p{\mathbf{p}}
\def\u{\mathbf{u}}
\def\j{\mathrm{j}}

\newtheorem{prop}{\textbf{Proposition}}

\begin{document}
\title{\LARGE Sparsification of Precoding Codebooks for PAPR Reduction via Grassmannian Representations}

\author{Joe~Asano,~\IEEEmembership{Student Member,~IEEE}, Yuto Hama,~\IEEEmembership{Member, IEEE}, Hiroki Iimori,~\IEEEmembership{Member, IEEE},\\ Szabolcs Malomsoky, and Naoki~Ishikawa,~\IEEEmembership{Senior~Member,~IEEE}.\thanks{J.~Asano and N.~Ishikawa are with the Faculty of Engineering, Yokohama National University, 240-8501 Kanagawa, Japan (e-mail: iskw@ieee.org). Y.~Hama, H.~Iimori, and S.~Malomsoky are with Ericsson Research, Ericsson Japan K. K., Yokohama, 220-0012 Kanagawa, Japan (e-mail: [yuto.hama, hiroki.iimori, szabolcs.malomsoky]@ericsson.com).
}}

\markboth{\today}
{Shell \MakeLowercase{\textit{et al.}}: Bare Demo of IEEEtran.cls for Journals}
\maketitle

\begin{abstract}
In this letter, we propose a sparsification method for precoding codebooks that reduces the peak-to-average power ratio (PAPR) while preserving the achievable rate. By exploiting the fact that precoder matrices lie on the Grassmann manifold, we formulate a codebook design problem that enables sparsification without modifying the existing feedback mechanism. We develop two sparsification approaches, namely exact sparsification via unitary transformation and approximate sparsification via sparse principal component analysis (SPCA), and integrate them into a unified design algorithm. The proposed sparsified codebooks incur negligible performance loss while reducing PAPR by more than $1~\text{dB}$ in uplink scenarios.
\end{abstract}


\IEEEpeerreviewmaketitle

\section{Introduction}
\label{sec:intro}
\IEEEPARstart{I}{N} the evolution toward 6G systems, uplink traffic is expected to increase significantly due to the growing number of connected devices~\cite{ericsson2025ericsson_short}. In uplink systems, power amplifier (PA) efficiency is inherently constrained at user equipment, and the high peak-to-average power ratio (PAPR) remains a major challenge in orthogonal frequency-division multiplexing (OFDM) signals, which are commonly used in modern wireless communication standards.
Extensive studies have characterized the statistical properties of PAPR and developed reduction techniques for OFDM and discrete Fourier transform spread OFDM (DFT-s-OFDM)~\cite{ochiai2001distribution_short, taojiang2008overview_short}, which are adopted in current 5G uplink systems. To further enhance both uplink capacity and coverage in beyond-5G and 6G systems, extending DFT-s-OFDM to support multi-stream transmission has attracted increasing interest, since the current 5G standard restricts it to single-stream transmission.

Spatial precoding plays an essential role in multiple-input multiple-output~(MIMO) systems.
In 5G uplink, codebook-based precoding is employed, where the codebooks composed of multiple precoders are standardized in~\cite{3gpp_ts_38211_v1820} for different numbers of antenna ports and data streams.
The precoding matrices can be interpreted as points on a Grassmann manifold~\cite{love2003grassmannian_short, love2005limited_short}, and under uncorrelated Rayleigh fading, codebook design is equivalent to subspace packing on the Grassmann manifold. However, existing codebooks adopt dense precoder structures designed to maximize array gain 
and their design does not take into account the use of multi-stream DFT-s-OFDM.
When applied to multi-stream transmission, such dense precoders break the single-carrier property of DFT-s-OFDM, resulting in increased PAPR. Recent work~\cite{asano2026low_short} addressed this issue by designing codebooks consisting exclusively of sparse precoders and demonstrated their effectiveness in reducing PAPR for multi-stream DFT-s-OFDM.

In this letter, we propose a sparsification method for precoding codebooks that reduces PAPR while maintaining the achievable rate. Unlike~\cite{asano2026low_short}, which designs new codebooks, the proposed method sparsifies existing codebooks standardized in~\cite{3gpp_ts_38211_v1820} while preserving the original codebook indices.
This approach enables effective PAPR reduction without affecting existing standard specifications, since dense precoders can be replaced with sparse precoders through a user-equipment-only implementation.
By exploiting the Grassmannian representation of precoder matrices, sparse precoders are constructed either via unitary transformations of the original precoders or via sparse principal component analysis (SPCA)~\cite{asteris2014sparse_short, zou2006sparse_short}. The main contributions are summarized as follows:
\begin{enumerate}
\item We propose a codebook-based precoding framework that uses sparse precoders instead of dense precoders for PAPR reduction in DFT-s-OFDM without performance degradation and without modifying the existing feedback mechanism.
\item We develop two closed-form sparsification methods on the Grassmann manifold, based on unitary transformations and SPCA.
\end{enumerate}

\section{System Model}
\label{sec:sys}
In this section, we describe the limited-feedback MIMO system~\cite{love2003grassmannian_short, love2005limited_short} with codebook-based precoding.

We consider a single-user uplink limited-feedback MIMO system with $N_t$ transmit and $N_r$ receive antennas, where $N_s \le \min(N_t,N_r)$ denotes the number of data streams. Let $\H_k \in \C^{N_r \times N_t}$ denote the channel matrix for the $k$-th subcarrier, $k=1,\ldots,K$. Let $\s_k \in \C^{N_s \times 1}$ denote the transmitted symbol vector on the $k$-th subcarrier, drawn from phase shift keying (PSK) or quadrature amplitude modulation (QAM) constellations and satisfying $\E[\s_k \s_k^\e]=\I_{N_s}$. The received signal $\y_k \in \C^{N_r \times 1}$ is given by
\begin{align}
    \y_k=\H_k\W_i\s_k+\mathbf{v}_k,
    \label{eq:sys}
\end{align}
where $\W_i \in \C^{N_t \times N_s}$ is selected from a finite codebook $\mathcal{W}={\W_1,\ldots,\W_{|\mathcal{W}|}}$, $\H_k$ is the channel matrix, and $\mathbf{v}_k \sim \CN(\mathbf{0},\sigma_v^2\I_{N_r})$ is additive white Gaussian noise.

We consider wideband precoding as in current 5G uplink systems, where a common precoder is applied across all subcarriers. Each precoder satisfies $\W_i^\e \W_i = \I_{N_s}$. With perfect channel state information at the receiver, 
the precoder is selected to maximize the achievable rate
\begin{align}
    R(\W_i)=\sum_{k=1}^K~ \log_2\det\left(\I_{N_s}+\frac{\rho}{N_s}\W_{i}^\e\H_k^\e\H_k\W_i \right),
    \label{eq:selection}
\end{align}
where $\rho=1/\sigma_v^2$ denotes the average signal-to-noise ratio (SNR).
As seen from~\eqref{eq:selection}, $R(\W_i)$ is invariant under right multiplication of $\W_i$ by any unitary matrix. Together with the orthonormality constraint, this implies that $\W_i$ can be regarded as a point on the Grassmann manifold $\G(N_t, N_s)$.
The receiver feeds back the index $\hat{i}$ corresponding to the optimal precoder as
\begin{align}
    \hat{i}=\underset{1\le i\le|\mathcal{W}|}{\text{argmax}}~R(\W_i),
    \label{eq:feedback}
\end{align}
using $\lceil \log_2|\mathcal{W}| \rceil$ bits.

\section{Proposed Sparsification Methods}
\label{sec:prop}
In this section, we describe a new problem formulation for reducing PAPR without modifying the existing feedback mechanism and the corresponding codebook design procedure.

\subsection{Problem Formulation}
To reduce PAPR in multi-stream DFT-s-OFDM, we propose a pre-mapped sparse codebook design that operates within the existing limited-feedback framework.
An overview of the system is summarized as follows:
\begin{itemize}
    \item The UE independently selects a precoder from the pre-mapped sparse codebook using the same feedback index as the original dense codebook.
    \item The selected precoder reduces PAPR with little performance loss in multi-stream DFT-s-OFDM due to its sparse structure.
\end{itemize}
In this system, the pre-mapped sparse codebook should preserve the performance and spatial characteristics (e.g., main lobe directions) of the original dense precoder at each index, since the UE independently selects the precoder without modifying the feedback mechanism. From the viewpoint of codebook design, each precoder represents a point on the Grassmann manifold $\G(N_t,N_s)$. Hence, the performance preservation problem can be interpreted as approximating each original precoder by a sparse precoder on the Grassmann manifold. On the Grassmann manifold, the distance between two points is defined as the chordal distance. Specifically, the chordal distance between two points $\W$ and $\P$, denoted by $d_{\mathrm{c}}(\W,\P)$, is defined as
\begin{align}
    d_{\mathrm{c}}(\W,\P)=\sqrt{N_s-\|\W^\e\P\|^2_\F}.
    \label{eq:chordal}
\end{align}
Let the original dense codebook be denoted as $\mathcal{W}=\{\W_1,\cdots,\W_{|\mathcal{W}|}\}$ and the pre-mapped sparse codebook as $\mathcal{P}=\{\P_1,\cdots,\P_{|\mathcal{P}|}\}$.
Based on the results demonstrated in~\cite{asano2026low_short}, we set the number of nonzero elements of each precoder matrix $\P_i$ to $N_t$, i.e., $\|\P_i\|_0=N_t$, taking into account the trade-off between sparsity and performance.
Since the optimization can be carried out independently for each index \( i \in \{1,\cdots,|\mathcal{W}|\} \), the pre-mapped sparse codebook design reduces to the following per-index problem:
\begin{align}
\begin{split}
    \underset{\mathbf{P}}{\operatorname{minimize}} \;& d_{\mathrm{c}}(\mathbf{W},\mathbf{P}) \\
    \text{s.t.}\;& \mathbf{P}\in\mathcal{G}(N_t,N_s),\; \|\mathbf{P}\|_0 = N_t.
\end{split}
\label{eq:obj_first}
\end{align}
Thus, the index \( i \) is omitted hereafter without loss of generality, since each precoder is optimized independently.

\subsection{Exact Sparsification via Unitary Transformation}
Regarding the problem defined in~\eqref{eq:obj_first}, when the optimal sparse precoder $\P$ exists, that is, the chordal distance between $\W$ and $\P$ is $0$, they span the same subspace, and there exists a unitary matrix $\U\in\C^{N_s\times N_s}$ such that
\begin{align}
    \P=\W\U.
    \label{eq:unitary_expression}
\end{align}
In this work, we demonstrate that the derivation of such a unitary matrix $\U$ and corresponding sparse precoder $\P$ can be obtained in closed form for a given dense precoder $\W$. 

Let $I_j\subset\{1,\cdots,N_t\}$ be the support set of $\p_j$,  i.e., $\text{supp}(\p_j)$, which represents the set of indices corresponding to the nonzero elements of each vector $\p_j$, where $\p_j\in\C^{N_t\times 1}$ denotes the $j$-th column vector of $\P$. 
Furthermore, we define $\mathcal{I}=\{I_1,\cdots, I_{N_s}\}$ as the collection of support sets $I_j$ for a given sparse matrix $\P$, that is, the set of the positions of all nonzero elements in the matrix. In this letter, we refer to $\mathcal{I}$ as a sparsity pattern. 
Since all sparse matrices $\P$ must be on the Grassmann manifold, $\P$ must satisfy the orthogonality constraint $\P^\e\P=\I_{N_s}$. Thus, the support sets for an arbitrary sparsity pattern $\mathcal{I}$ must be mutually disjoint, i.e.,
$I_j\cap I_k=\varnothing$ for all $j,k\in\{1,\cdots, N_s\}$. When we let $n(N_t,N_s)$ denote the number of candidates for sparsity patterns that satisfy this constraint for a given matrix size $N_t\times N_s$, $n(N_t,N_s)$ can be simply derived as
\begin{align}
    n(N_t,N_s)=\frac{1}{N_s!}\sum_{m=0}^{N_s}(-1)^m\binom{N_s}{m}(N_s-m)^{N_t}.
    \label{eq:n(Nt,Ns)}
\end{align}
This corresponds to the Stirling number of the second kind, which gives the total number of ways to partition $N_t$ distinct elements into $N_s$ nonempty subsets. 
For instance, when the matrix size is given by $(N_t,N_s)=(4,2)$, the number of candidates is obtained as $n(4,2)=7$, and the resulting sparsity patterns are denoted as
$\mathcal{I}_1=\{\{1,2\},\{3,4\}\},~\mathcal{I}_2=\{\{1,3\},\{2,4\}\}, $ $\mathcal{I}_3=\{\{1,4\},\{2,3\}\}, ~\mathcal{I}_4=\{\{1\},\{2,3,4\}\},$ $\mathcal{I}_5=\{\{1,3,4\}\{2\}\},$ $\mathcal{I}_6=\{\{1,2,4\},\{3\}\},$ and $\mathcal{I}_7=\{\{1,2,3\},\{4\}\}$.

For a given sparsity pattern $\mathcal{I}=\{I_1,\cdots,I_{N_s}\}$ defined in~\eqref{eq:n(Nt,Ns)}, let $J_j$ denote the complement of each support set $I_j$, i.e., $J_j=I_j^{c}$. For each column vector $\p_j$, the entries corresponding to the indices $J_j$ must be zero, i.e., $\p_j[J_j]=\mathbf{0}\in\C^{|J_j|\times 1}$. When the unitary matrix representation defined in~\eqref{eq:unitary_expression} holds, the constraint
\begin{align}
    (\W\u_j)[J_j]=\mathbf{0},
    \label{eq:const}
\end{align}
is satisfied, where $\u_j$ is the $j$-th column vector of $\U$. The constraint~\eqref{eq:const} can be equivalently rewritten as
\begin{align}
    \u_j\in\text{ker}(\W[J_j, :]),
    \label{eq:const_ker}
\end{align}
where $\text{ker}(\cdot)$ denotes the null space of a matrix. Regarding~\eqref{eq:const_ker}, a feasible unitary vector $\u_j$ exists when the null space of $\W[J_j,:]$ has a nonzero dimension.
Additionally, since $\U$ is a unitary matrix, 
the following orthonormality constraints must hold for any pair of column indices $1\le j<k\le N_s$:
\begin{align}
    \begin{split}
        \u_j^\e\u_k=0, ~
        \|\u_j\|=1.
    \end{split}
\label{eq:const_unitary}
\end{align}
To make the above conditions explicit and verifiable, we state the following proposition.

\begin{prop}
\label{prop:feasibility}
    For a given dense precoder $\W\in\G(N_t,N_s)$ and a sparsity pattern $\mathcal{I}=\{I_1,\cdots, I_{N_s}\}$, there exists a unitary matrix $\U=[\u_1,\cdots, \u_{N_s}]\in\C^{N_s\times N_s}$ satisfying~\eqref{eq:unitary_expression} if and only if one can sequentially construct vectors $\u_1,\cdots, \u_{N_s}$ such that
    \begin{align}
        \u_j\in\text{ker}(\W[J_j, :]),~\u_j^\e\u_k=0~(j\neq k),~\|\u_j\|=1,
        \label{eq:prop}
    \end{align}
    for all $j,k\in\{1,\cdots, N_s\}$. Equivalently, the sparsity pattern $\mathcal{I}$ is feasible if and only if, at each step $j$, there exists a nonzero vector in $\text{ker}(\W[J_j, :])\bigcap\text{span}\{\u_1,\cdots ,\u_{j-1}\}^{\perp}$.
\end{prop}

Based on Proposition~\ref{prop:feasibility}, the feasibility of a given sparsity pattern can be verified constructively. When the above condition is satisfied, the desired unitary matrix $\U$ can be obtained in closed form by the following procedure. Let $\Q_j\in\C^{N_s\times N_s}$ denote the projection matrix onto the null space of $\W[J_j,:]$, which is defined as
\begin{align}
    \Q_j=\I_{N_s}-\W[J_j,:]^\dagger\W[J_j,:],
    \label{eq;projection}
\end{align}
where $(\cdot)^\dagger$ denotes the pseudoinverse. For an arbitrary column vector $\x_j\in\C^{N_s}$, the vector $\Q_j\x_j$ must lie in the null space of $\W[J_j,:]$, i.e., $\ker(\W[J_j,:])$. 
Next, to satisfy the orthonormality condition given by~\eqref{eq:const_unitary}, we perform the Gram-Schmidt orthogonalization process, which is expressed as
\begin{align}
    \u_j = \frac{\left(\I_{N_s}-\sum_{m=1}^{j-1}\u_m\u_m^\e \right)\Q_j\x_j }{\left\|\left(\I_{N_s}-\sum_{m=1}^{j-1}\u_m\u_m^\e \right)\Q_j\x_j\right\|}.
    \label{eq:gram-schmidt}
\end{align}
By iteratively applying this procedure for $j=1,\cdots,N_s$, the desired unitary matrix $\U=[\u_1,\cdots ,\u_{N_s}]$ is obtained, when the feasibility condition in Proposition~\ref{prop:feasibility} is satisfied. In this case, each step yields a nonzero vector, and thus the Gram-Schmidt procedure does not degenerate.

For example, we consider the sparsification of the original dense precoder $\W$, which is denoted as
\begin{align}
    \W=\frac{1}{2} \begin{bmatrix}
        1 & -\j & 1 & -\j \\
        1 & -\j & -1 & \j 
    \end{bmatrix}^\mathrm{T}.
    \label{eq:example}
\end{align}
When the sparsity pattern $\mathcal{I}=\{I_1,I_2\}$ is given as $I_1=\{1,2\}$ and $I_2=\{3,4\}$, their complement subsets are denoted as $J_1=\{3,4\}$ and $J_2=\{1,2\}$ respectively. Thus, the submatrices corresponding to these complement subsets are denoted as
\begin{align}
    \W[J_1,:]=\frac{1}{2}\begin{bmatrix}
        1 &-1 \\
        -\j & \j
    \end{bmatrix}, ~
    \W[J_2,:]=\frac{1}{2}\begin{bmatrix}
        1 & 1 \\
        -\j & -\j
    \end{bmatrix},
    \label{eq:complement}
\end{align}
respectively. Here, each null space is also denoted as 
\begin{align}
    \begin{split}
        &\ker\left(\W[J_1,:] \right)=\mathrm{span}\left([1,1]^\mathrm{T}\right), \\
        &\ker\left(\W[J_2,:] \right)=\mathrm{span}\left([1,-1]^\mathrm{T}\right).
    \end{split}
    \label{eq:null}
\end{align}
Since these null spaces are mutually orthogonal, we automatically obtain the orthonormal column vectors $\u_1$ and $\u_2$ as
$\u_1=1/\sqrt{2}~[1,1]^\mathrm{T}$ and $\u_2=1/\sqrt{2}~[1,-1]^\mathrm{T}$ respectively.
Finally, the desired unitary matrix $\U$ is obtained as $\U=[\u_1,\u_2]$, and the corresponding sparse precoder $\P$ is also obtained as
\begin{align}
    \P=\W\U=\frac{1}{\sqrt{2}}\begin{bmatrix}
        1 & -\j & 0 & 0 \\
        0 & 0 & 1 & -\j
    \end{bmatrix}^\mathrm{T}.
    \label{eq:P}
\end{align}
At this point, $\P$ and $\W$ represent the same subspace on the Grassmann manifold, and the resulting chordal distance between these two points is $0$.

\subsection{Approximate Sparsification via SPCA} 
In this subsection, we describe a process for approximating the sparsification problem in~\eqref{eq:obj_first} based on SPCA when no unitary matrix $\U$ exists for the problem defined in~\eqref{eq:obj_first}.
For a given dense matrix $\W$, minimizing the chordal distance to $\P$ can be expressed as the following maximization problem based on the inner product of $\W$ and $\P$:
\begin{align}
    \underset{\P}{\operatorname{maximize}}~\|\W^\e\P\|_\F^2.
    \label{eq:obj_innner}
\end{align}
This formulation corresponds to searching for a sparse matrix $\P$ that maximizes the inner product with the original dense matrix $\W$, and this problem can be formulated as SPCA, a well-known problem in the literature~\cite{asteris2014sparse_short, zou2006sparse_short}. In this sense, the sparsification of a point on the Grassmann manifold can be interpreted as an SPCA problem. According to the SPCA formulation, \eqref{eq:obj_innner} can be rewritten as
\begin{align}
    \underset{\P}{\operatorname{maximize}}~\sum_{j=1}^{N_s}\p_j^\e\W\W^\e\p_j.
    \label{eq:obj_SPCA}
\end{align}

Regarding~\eqref{eq:obj_SPCA}, we consider the problem of maximizing the inner product $\p_j^\e\W\W^\e\p_j$ for each column vector $\p_j$, subject to the constraints $\text{supp}(\p_j)=I_j\in\mathcal{I}$ and $\|\p_j\|=1$.
For a given support set $I_j$, the optimal nonzero entries of $\p_j$ can be obtained in closed form. Specifically, this problem reduces to the Rayleigh quotient maximization problem.
Here, let $\A_j=(\W\W^\e)[I_j,I_j]\in\C^{|I_j|\times |I_j|}$ denote the submatrix corresponding to the rows and columns indexed by $I_j$, and let $\b_j=\p_j[I_j]\in\C^{|I_j|\times1}$ denote the subvector of $\p_j$ restricted to the support set $I_j$.
Then, the optimal subvector $\hat{\b_j}$ that maximizes the inner product $\b_j^\e\A_j\b_j$ is given by the eigenvector associated with the largest eigenvalue of $\A_j$, denoted by $\lambda_{\text{max}}(\A_j)$, from the Rayleigh quotient maximization result. Hence, the corresponding optimal vector $\hat{\p_j}$ that maximizes the inner product $\p_j^\e\W\W^\e\p_j$ is obtained by embedding the subvector $\hat{\b_j}$ into the indices $I_j$ and setting all remaining elements to zero. 

Consequently, the SPCA optimization problem defined in~\eqref{eq:obj_SPCA} can be rewritten as
\begin{align}
    \underset{\mathcal{I}_k}{\operatorname{argmax}}~\sum_{j=1}^{N_s}\lambda_{\text{max}}(\A_j),~\forall k\in\{1,\cdots, n(N_t,N_s)\} .
    \label{eq:obj_comb}
\end{align}
That is, the problem is formulated as a combinatorial optimization problem of selecting the optimal sparsity pattern $\mathcal{I}$ from $n(N_t,N_s)$ candidates. Finally, the optimal matrix $\hat{\P}$ is given by $\hat{\P}=[\hat{\p_1},\cdots,\hat{\p_{N_s}}]$, and the resulting optimal codebook $\mathcal{P}$ is also obtained by performing this optimization independently for a given codebook size $|\mathcal{P}|$.

\subsection{Overall Sparse Codebook Construction Algorithm}

\begin{algorithm}[t]
\caption{Proposed Sparse Codebook Construction}
\label{alg:prop}
\begin{algorithmic}[1]
    \STATE \textbf{Input:} Dense codebook $\mathcal{W}=\{\W_1,\cdots,\W_{|\mathcal{W}|}\}$, candidate sparsity patterns $\{\mathcal{I}_1,\cdots,\mathcal{I}_{n(N_t,N_s)}\}$

    \FOR{each codebook index $i=1$ to $|\mathcal{W}|$}
        \STATE $b_{\text{feasible}}$ $\leftarrow$ false

        \FOR{each sparsity pattern $\mathcal{I}_k \in \{\mathcal{I}_1,\cdots,\mathcal{I}_{n(N_t,N_s)}\}$}
            \STATE Attempt to construct unitary matrix $\U_i$ via~\eqref{eq:gram-schmidt}.
            \IF{the constructed $\U_i$ satisfies the condition~\eqref{eq:prop} in Proposition~\ref{prop:feasibility}}
                \STATE $\P_i \leftarrow \W_i \U_i$
                \STATE $b_{\text{feasible}}$ $\leftarrow$ true 
                \STATE \textbf{break}
            \ENDIF
        \ENDFOR
        \IF{$b_{\text{feasible}}$ = false}
            \FOR{each sparsity pattern $\mathcal{I}_k$}
                \STATE Solve the SPCA problem via~\eqref{eq:obj_comb}.
            \ENDFOR
            \STATE Select the optimal sparsity pattern.
            \STATE Construct $\P_i$ based on the SPCA solution.
        \ENDIF
    \ENDFOR

    \STATE \textbf{Output:} Sparse codebook $\mathcal{P}=\{\P_1,\cdots,\P_{|\mathcal{P}|}\}$
\end{algorithmic}
\end{algorithm}

We summarize the proposed sparse codebook design method as follows. For each precoder in the original dense codebook, candidate sparsity patterns are generated via~\eqref{eq:n(Nt,Ns)} based on the combinatorial structure for the given configuration $(N_t,N_s)$. For each sparsity pattern, we first attempt to construct a unitary matrix $\U$ based on the procedure in~\eqref{eq:gram-schmidt}. In particular, this step aims to sequentially construct an orthonormal set of vectors satisfying the conditions in~\eqref{eq:prop} of Proposition~\ref{prop:feasibility}. The construction succeeds if and only if the given sparsity pattern is feasible, i.e., when exact sparsification is possible under the corresponding sparsity constraints.
If a feasible unitary matrix exists, the sparse precoder is obtained as $\P~=\W\U$ using the closed-form solution, which preserves the subspace exactly. Otherwise, if no feasible unitary matrix is found for all candidate sparsity patterns, we instead compute an approximate solution by solving the SPCA problem for each candidate sparsity pattern. The sparsity pattern that maximizes~\eqref{eq:obj_comb} is selected, and the corresponding sparse precoder is obtained accordingly. The detailed procedure of the proposed sparse codebook construction is summarized in Algorithm~\ref{alg:prop}.

\section{Numerical Results}
\label{sec:res}
In this section, we evaluate the performance of the proposed sparsification method in terms of achievable rate, PAPR, and sparsification distortion with respect to matrix dimensions.

\subsection{Achievable Rate}
\begin{figure}[tb]
    \centering
    \includegraphics[width=\linewidth]{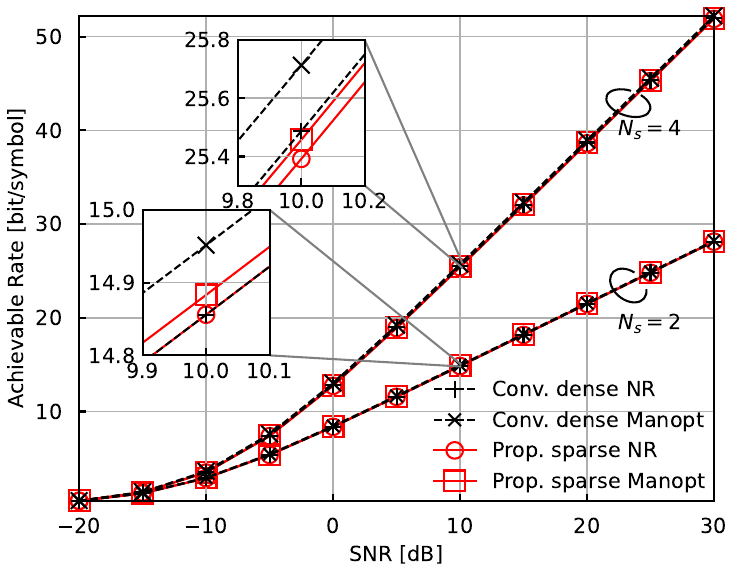}
    \caption{Achievable rate under uncorrelated Rayleigh fading channels with $(N_t,N_s)=(4,2)$ and $(8,4)$.}
    \label{fig:achievable_rate}
\end{figure}

First, Fig.~\ref{fig:achievable_rate} shows the achievable rate under uncorrelated Rayleigh fading channels. 
The original dense precoding codebooks are 
the 5G NR codebooks specified in~\cite{3gpp_ts_38211_v1820}, and the codebooks optimized to maximize the minimum chordal distance~\cite{love2003grassmannian_short, love2005limited_short}, which we refer to as Manopt.
Based on the NR specifications, we consider $(N_t,N_s,|\mathcal{W}|)=(4,2,8)$ and $(8,4,24)$. 
Since we only need to evaluate the performance degradation caused by sparsification in this work, we adopted the indices corresponding to the dense matrices from Table 6.3.1.5-5 in~\cite{3gpp_ts_38211_v1820} for $(N_t,N_s)=(4,2)$. 
In both cases, the number of receive antennas was set to $N_r=32$, and the channel matrix is assumed as uncorrelated Rayleigh fading channels, whose entries follow $\CN(0,1)$ respectively.
As described in Section~\ref{sec:prop}, the sparse precoder is evaluated using the same index selected from the original dense codebooks.

As shown in Fig.~\ref{fig:achievable_rate}, for $(N_t,N_s)=(4,2)$, the proposed sparsified codebook achieved no performance loss with the NR codebook, since the unitary transformation succeeded for all indices. For Manopt inputs, although the unitary transformation was infeasible and SPCA-based approximation was applied, the SNR loss was limited to approximately $0.1~\text{dB}$. For $(N_t,N_s)=(8,4)$, with NR inputs, the unitary transformation succeeded for some indices, while SPCA was applied to others, resulting in a slight SNR loss of about $0.05~\text{dB}$. For Manopt inputs, the sparsified codebook was obtained via SPCA, with an SNR loss of approximately $0.2~\text{dB}$.

\subsection{PAPR}
\begin{figure}[tb]
    \centering
    \includegraphics[width=\linewidth]{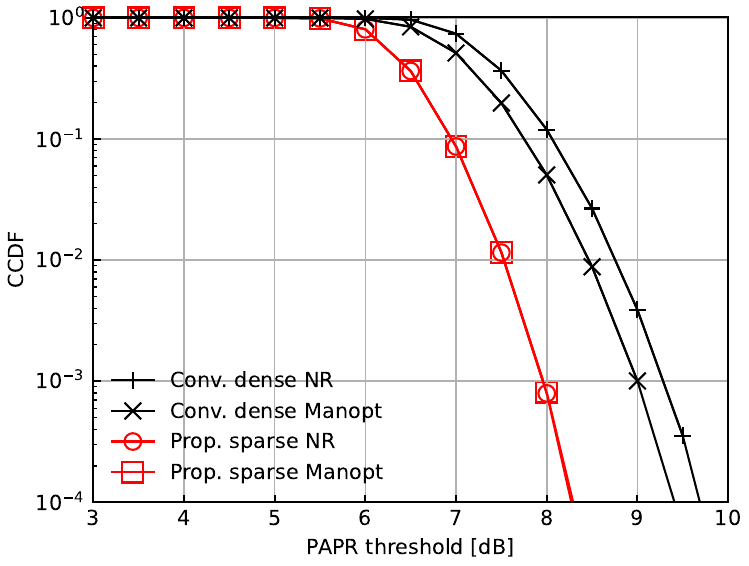}
    \caption{Comparison of PAPR performance in DFT-s-OFDM with $(N_t,N_s)=(4,2)$, a codebook size of $8$, $52$ PRBs (624 subcarriers), $1024$-point FFT, and $8$ times oversampling.}
    \label{fig:PAPR}
\end{figure}

Next, we compare the PAPR of DFT-s-OFDM using the complementary cumulative distribution function (CCDF). The PAPR is computed from the time-domain signal after codebook-based precoding is applied to each subcarrier of the DFT-precoded sequences, and then averaged over each antenna. For accurate PAPR evaluation, we perform the oversampling for the time domain signals. The detailed procedure is described in~\cite{berardinelli2008ofdma_short, hama2023timefrequency_short}.

The simulations assume 52 physical resource blocks (PRBs), i.e., 624 subcarriers with an FFT size of 1024, and the input symbols modulated by 4-QAM. To ensure accurate PAPR estimation, the time-domain signals are generated with 8 times oversampling. The same codebooks for $(N_t,N_s)=(4,2)$ are used as in the achievable rate evaluation.

As shown in Fig.~\ref{fig:PAPR}, the proposed sparsified codebooks achieved a PAPR reduction of more than $1~\text{dB}$ at $\text{CCDF}=10^{-2}$ for both NR and Manopt inputs. 
This PAPR reduction enables higher transmit power due to lower required PA backoff, leading to a potential improvement in achievable rate under the same power constraint.

\subsection{Sparsification Distortion}
\begin{figure}[tb]
    \centering
    \subfloat[Fixed $N_s$ with varying $N_t$.]{
        \includegraphics[width=\linewidth]{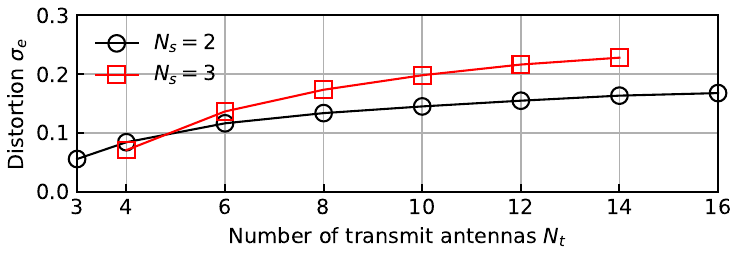}
    } \\
    \subfloat[Fixed $N_t$ with varying $N_s$.]{
        \includegraphics[width=\linewidth]{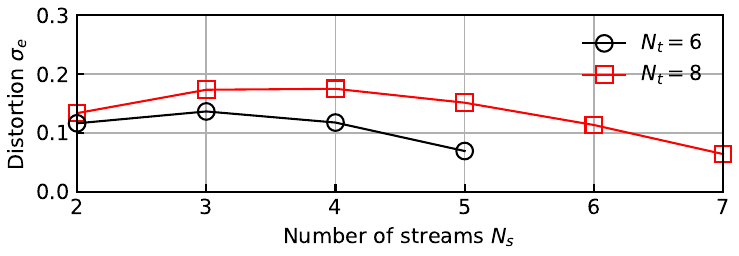}
    }
    \caption{Sparsification distortion comparisons when varying the matrix size.}
    \label{fig:distortion}
\end{figure}

Finally, we evaluate the sparsification distortion of the proposed SPCA-based approximation method using randomly generated matrices on the Grassmann manifold $\G(N_t,N_s)$. For fair comparison across different $(N_t,N_s)$, the distortion is defined as the chordal distance-based normalized mean square error. For the dense precoder $\W$ and sparsified precoder $\P$, the distortion $\sigma_e$ is given by
\begin{align}
    \sigma_e=\mathbb{E}\left[\frac{\|\mathbf{P}\mathbf{P}^{\mathrm{H}}-\mathbf{W}\mathbf{W}^{\mathrm{H}}\|_{\mathrm{F}}^2}{2\|\mathbf{W}\mathbf{W}^{\mathrm{H}}\|_{\mathrm{F}}^2}\right]=\mathbb{E}\left[\frac{\|\mathbf{P}\mathbf{P}^{\mathrm{H}}-\mathbf{W}\mathbf{W}^{\mathrm{H}}\|_{\mathrm{F}}^2}{2N_s}\right].
\label{eq:distortion}
\end{align}

Fig.~\ref{fig:distortion}(a) shows the sparsification distortion for fixed $N_s=2,3$ with varying $N_t$. The distortion increased monotonically with $N_t$. Since $\W\in\C^{N_t\times N_s}$ lies on the Grassmann manifold of dimension $N_s(N_t-N_s)$, the manifold dimension increases with $N_t$. However, the number of nonzero elements was fixed at $N_t$, limiting the effective degrees of freedom relative to the manifold dimension, which led to increased distortion under the sparsity constraint.
Fig.~\ref{fig:distortion}(b) shows the distortion for fixed $N_t=6,8$ with varying $N_s$. The distortion was concave in $N_s$ and was maximized at $N_s=N_t/2$, consistent with the concave quadratic form of $N_s(N_t-N_s)$.

In all cases, the distortion remained around $0.2$ and resulted in negligible performance degradation, as shown in Fig.~\ref{fig:achievable_rate}.

\section{Conclusion}
\label{sec:conc}
In this letter, we proposed a sparsification method for precoding codebooks that preserves the achievable rate while reducing PAPR. By exploiting the Grassmannian property of the precoder matrices, we formulated a codebook design problem that enables sparsification without modifying the existing feedback mechanism. Two methods were developed, namely exact sparsification via unitary transformation and SPCA-based approximation, within a unified design algorithm. The proposed codebooks incurred negligible performance loss while reducing PAPR in DFT-s-OFDM systems. 

\footnotesize{
	\bibliographystyle{IEEEtran}
	\bibliography{main}

@article{3gpp_ts_38211_v1820,
  title = {{{TS}} 138 211 - {{V18}}.2.0 - {{5G}}; {{NR}}; {{Physical}} Channels and Modulation ({{3GPP TS}} 38.211 Version 18.2.0 {{Release}} 18)},
  author = {{3GPP}},
  year = {2024}
}

@article{ericsson2025ericsson_short,
  author = {{Ericsson}},
  title = {Ericsson Mobility Report},
  journal = {Ericsson},
  year = {2025}
}

@article{ochiai2001distribution_short,
  author = {Ochiai, H. and Imai, H.},
  title = {On the Distribution of the Peak-to-Average Power Ratio in {OFDM} Signals},
  journal = {IEEE Trans. Commun.},
  volume = {49},
  number = {2},
  pages = {282--289},
  year = {2001}
}

@article{taojiang2008overview_short,
  author = {Jiang, T. and Wu, Y.},
  title = {An Overview: Peak-to-Average Power Ratio Reduction Techniques for {OFDM} Signals},
  journal = {IEEE Trans. Broadcast.},
  volume = {54},
  number = {2},
  pages = {257--268},
  year = {2008}
}

@article{love2003grassmannian_short,
  author = {Love, D. and Heath, R. and Strohmer, T.},
  title = {GGrassmannian Beamforming for Multiple-Input Multiple-Output Wireless Systems},
  journal = {IEEE Trans. Inf. Theory},
  volume = {49},
  number = {10},
  pages = {2735--2747},
  year = {2003}
}

@article{love2005limited_short,
  author = {Love, D. and Heath, R.},
  title = {Limited Feedback Unitary Precoding for Spatial Multiplexing Systems},
  journal = {IEEE Trans. Inf. Theory},
  volume = {51},
  number = {8},
  pages = {2967--2976},
  year = {2005}
}

@article{asano2026low_short,
  author = {Asano, J. and others},
  title = {Low-Complexity and Power Efficient Precoding Codebook Design on Sparse {Grassmannian}},
  journal = {arXiv preprint},
  note = {arXiv:2603.02515},
  year = {2026}
}

@article{asteris2014sparse_short,
  author = {Asteris, M. and others},
  title = {The Sparse Principal Component of a Constant-Rank Matrix},
  journal = {IEEE Trans. Inf. Theory},
  volume = {60},
  number = {4},
  pages = {2281--2290},
  year = {2014}
}

@article{berardinelli2008ofdma_short,
  author = {Berardinelli, G. and others},
  title = {{OFDMA} vs. {SC-FDMA}: Performance Comparison in Local Area {IMT-A} Scenarios},
  journal = {IEEE Wireless Commun.},
  volume = {15},
  number = {5},
  pages = {64--72},
  year = {2008}
}

@article{hama2023timefrequency_short,
  author = {Hama, Y. and Ochiai, H.},
  title = {Time-Frequency Domain Non-Orthogonal Multiple Access for Power Efficient Communications},
  journal = {IEEE Trans. Wireless Commun.},
  volume = {22},
  number = {9},
  pages = {5711--5724},
  year = {2023}
}

@article{zou2006sparse_short,
  author = {Zou, H. and Hastie, T. and Tibshirani, R.},
  title = {Sparse Principal Component Analysis},
  journal = {J. Comput. Graph. Statist.},
  volume = {15},
  number = {2},
  pages = {265--286},
  year = {2006}
}
}

\end{document}